\documentclass[12pt]{iopart}

\usepackage{iopams}

\begin{document}

\title[The octonionic eigenvalue problem]{The octonionic eigenvalue problem}

\author{Stefano De Leo$^1$ and Gisele Ducati$^2$}

\address{$^1$ Department of Applied Mathematics, Universidade Estadual de Campinas}
\address{$^2$ Center of Mathematics, Computation and Cognition,
Universidade Federal do ABC}

\ead{deleo@ime.unicamp.br and ducati@ufabc.edu.br\\ {\sc Journal of Physics A {\bf 45}, 315203-17 (2012)} }

\begin{abstract}
By using a real matrix translation, we propose
a coupled eigenvalue problem for octonionic operators. In view of possible
applications in quantum mechanics, we also discuss the hermiticity of
such operators. Previous  difficulties in
formulating a consistent octonionic Hilbert space are solved by
using the new coupled eigenvalue problem and introducing an appropriate scalar product
for the probability amplitudes.
\end{abstract}

\pacs{02.30.Tb, 03.65.Fd}

\maketitle


\section{Introduction}

Developments in quaternionic matrix theory produced
interesting and important results in approaching and solving
quaternionic eigenvalue problems\,\cite{DS00,DSS02} and linear
differential equations with quaternionic
coefficients\,\cite{DD01,DD03}. This renewed the interest in
studying quaternionic formulations of quantum mechanics\,\cite{ADL}.
Previous discussions on quaternionic diffusion and tunnelling
phenomena\,\cite{P79,K84,K88,DM89,DM92,DDN01} and recent analysis of
confined states\,\cite{DD05} are now based on a more solid
mathematical understanding of the quaternionic structures involved
in such physical problems. Consequently, many of the previous
hidden aspects of the theory have been clarified and more
convincing proposals of quaternionic deviations from complex
theory can be now formulated. If quaternions, due to their
non-commutativity, represent a challenge for mathematicians and
physicists, the use of octonions to formulate quantum theories,
due to their non-commutativity and non-associativity, seems a very
hard challenge. It soon becomes clear that describing the physical
world in terms of octonionic mathematical structures involves many
conceptual (algebraic and analytic) difficulties that the lack of
associativity inevitably conjures up\,\cite{D94,GT96}. In this
spirit, the paper was intended as an attempt to motivate and
stimulate the study of octonionic mathematical problems in view of
possible applications in physical theory. The main difficulty in
carrying out octoninic formulations of quantum mechanics is
related to the appropriate definition of octonionic Hilbert spaces
and scalar products\,\cite{ADL,DK96a,DK96b}. In the next section, we
shall come back to this point and present a detailed discussion on
octonionic operators and complex geometry. We shall see that the
use of basic concepts in quantum mechanics, such as the correct
choice of the number system to appropriately define amplitudes of
probabilities, suggests the introduction of coupled octonionic
eigenvalue equations where two real parameters play the role that
complex eigenvalues do in complex and quaternionic quantum
mechanics.

Before to proceed with the main topic of our paper, i.e. the
octonionic eigenvalue problem in quantum mechanics, let us set up
notation and terminology, and study a simple eigenvalue problem by
using complex, quaternionic and octonionic algebras. This shall
elucidate in a practical way some of the difficulties due to the
lost of commutativity and associativity.

Let $\mathbb{O}$ be the octonionic division algebra. A generic
element in this algebra will be represented by
\begin{equation}
o = r_{0} + \displaystyle \sum_{m=1}^{7} r_m
e_m~,~~~r_{0,1,\cdots,7} \in \mathbb{R}\,\,,
\end{equation}
where $e_m$ are the octonionic immaginary units obeying the
following  non-commutative and non-associative algebraic rules
\begin{equation}
e_m e_n =- \, \delta_{mn} + \epsilon_{mnp}\, e_p\,\,,
\end{equation}
with {\footnotesize $m,n,p=1,2,\cdots,7$} and $\epsilon_{mnp}$
which is a  totally antisymmetric tensor equal to the unit for the
seven (quaternionic) combinations
\begin{equation}
123,~145,~176,~246,~257,~347~~~\mbox{and}~~~365\, \, .
\end{equation}
 The conjugate, the norm and the inverse of an octonion  are respectively
defined by
\begin{eqnarray}
o^{\dag} &= &r_0 - \sum_{m=1}^{7} r_m e_m\,\,,\nonumber \\
 N(o) & = & \sqrt{o^{\dag} o}
= \sqrt{o\,o^\dag} = \sqrt{r_{0}^{2} + \cdots + r_{7}^{2}}\,\,,\nonumber \\
\nonumber \\  o^{-1} &= &o^{\dag}/N(o)~~~~(0 \neq 0)\, \,.
\end{eqnarray}
Let us now analyze a particular eigenvalue problem in four different contexts. Given the
Hermitian matrix
\begin{equation}
M = \left(  \begin{array}{rr}
          1\,\, & \,\,\,\,\,e_{1}\,\\
          - \,e_{1} & 1\,\,\,
        \end{array}  \right)\, \, ,
\end{equation}
we aim to find its complex, quaternionic and octonionic
eigenvalues. In this paper, $e_{1}$ represents  the imaginary
unit of the complex field, $\mathbb{C}$, and $e_{1}, e_{2}$ and
$e_{3}$ the imaginary units of the quaternionic field,
$\mathbb{H}$.\\

{\bf (i)} The complex eigenvalue problem ({\sc CEP}),
\begin{equation}
\label{cep} \mbox{{\sc CEP}}:\hspace*{.5cm}
 M\,\Psi = \lambda \,
\Psi\,\, ,\,\,\,\,\, \Psi=\left(
\begin{array}{c} \psi_{a}\\\psi_{b} \end{array}\right) \,\,,
\,\,\,\,\,\psi_{a,b}\,\,\, \mbox{and} \,\,\, \lambda \in
\mathbb{C}\,\,,
\end{equation}
can be easily solved by standard calculations. Its eigenvectors
and eigenvalues are
\[
\left\{\, \psi_{a}\, ,\, \psi_{b}\,;\,\lambda \,\right\}_{1,2}=
\left\{\, z\, ,\, e_{1}\, z\,;\,0
\,\right\}_{1}\,\,\,\mbox{and}\,\,\, \left\{\, z\, ,\, -\,
e_{1}\, z\,;\,2 \,\right\}_{2}\,\,,\,\,\,\,\, z\in
\mathbb{C}\,\,.\]

\noindent{\bf (ii)} In exactly the same way, we can
establish the quaternionic version of Eq.(\ref{cep}). Due to the
non-commutativity of quaternions we have to introduce left and
right eigenvalue problems. Let us first discuss the quaternionic
left eigenvalue problem ({\sc QlEP}),
\begin{equation}
\label{qlep} \mbox{{\sc QlEP}}:\hspace*{.5cm} M\,\Psi = \lambda \,
\Psi\,\, ,\,\,\,\,\, \Psi=\left(
\begin{array}{c} \psi_{a}\\\psi_{b} \end{array}\right) \,\,,
\,\,\,\,\,\psi_{a,b}\,\,\, \mbox{and} \,\,\, \lambda \in
\mathbb{H}\,\,.
\end{equation}
After some algebraic manipulations, we find the following
eigenvectors and eigenvalues
\[
\left\{\, \psi_{a}\, ,\, \psi_{b}\,;\,\lambda \,\right\}=
\left\{\, q\, ,\, (\, \alpha \,e_{1} + \beta \,e_{2}+\gamma
\,e_{3}\,)\,q \,;\,1-\alpha +\beta \,e_{3} -\gamma \,e_{2}
\,\right\}\,\,,\]
where
\[\alpha^{2}+\beta^{2}+\gamma^{2}=1\,\,,\hspace*{.5cm}
\alpha,\beta,\gamma \in \mathbb{R}\hspace*{.5cm}\mbox{and}
\hspace*{.5cm} q\in \mathbb{H}\,.\] In
addition to the complex result ($\beta=\gamma=0$ and $q=z$) we
find a surprising result, that is the Hermitian matrix $M$ has
{\em quaternionic left} eigenvalues. This apparent paradox is soon
explained by observing that, due to the non-commutativity of the
quaternionic algebra, the standard proof used to show that complex
Hermitian matrices have {\em real} eigenvalues,
\[ \Psi^{\dag}(M \Psi)= (M \Psi)^{\dag}\Psi \,\,\,\Rightarrow \,\,\, \Psi^{\dag}\lambda
\Psi = (\lambda \Psi)^{\dag} \Psi \, [=\Psi^{\dag}\lambda^{\dag}
\Psi]
\]
fails for quaternions. In fact, due to the position of the
quaternionic eigenvalue $\lambda$,
\[ \Psi^{\dag}\lambda \Psi =\Psi^{\dag}\lambda^{\dag} \Psi \,\,\,
\Rightarrow \hspace*{-.4cm}/\,\,\,\,\,\,\,
\lambda=\lambda^{\dag}\,\,.\]

\noindent{\bf (iii)} To overcome this
difficulty, we change the position of the eigenvalue in
Eq.(\ref{qlep}), and introduce the quaternionic right eigenvalues
problem ({\sc QrEP}),
\begin{equation}
\label{qrep}\mbox{{\sc QrEP}}:\hspace*{.5cm} M\,\Psi =
\Psi\,\lambda\,\, ,\,\,\,\,\, \Psi=\left(
\begin{array}{c} \psi_{a}\\\psi_{b} \end{array}\right) \,\,,
\,\,\,\,\,\psi_{a,b}\,\,\, \mbox{and} \,\,\, \lambda \in
\mathbb{H}\,\,.
\end{equation}
Simple calculations show that the {\sc QrEP} has the same
eigenvalues of the {\sc CEP}, i.e.
\[
\left\{\, \psi_{a}\, ,\, \psi_{b}\,;\,\lambda \,\right\}_{1,2}=
\left\{\, q\, ,\, e_{1}\, q\,;\,0
\,\right\}_{1}\,\,\,\mbox{and}\,\,\, \left\{\, q\, ,\, -\,
e_{1}\, q\,;\,2 \,\right\}_{2}\,\,.\] The position of the
quaternionic eigenvalue plays a fundamental role to guarantee real
eigenvalues for   Hermitian matrices,
\[ \Psi^{\dag}(M \Psi)= (M \Psi)^{\dag}\Psi \,\,\,\Rightarrow  \,\,\,
\Psi^{\dag} \Psi \,\lambda = (\Psi \,\lambda)^{\dag} \Psi \,
[=\lambda^{\dag} \,\Psi^{\dag}\Psi] \,\,\,\Rightarrow\,\,\,
\lambda=\lambda^{\dag}\,\, .\]

\noindent{\bf (iv)} What happens for
quaternions suggests to consider an octonionic right-eigenvalue
problem ({\sc OrEP}),
\begin{equation}
\label{orep} \mbox{{\sc OrEP}}:\hspace*{.5cm} M\,\Psi =
\Psi\,\lambda\,\, ,\,\,\,\,\, \Psi=\left(
\begin{array}{c} \psi_{a}\\\psi_{b} \end{array}\right) \,\,,
\,\,\,\,\,\psi_{a,b}\,\,\, \mbox{and} \,\,\, \lambda \in
\mathbb{O}\,\,.
\end{equation}
The matrix equation (\ref{orep}) implies
\[ - \,\psi_b^{-1}\,(\,e_{1}\psi_a\,)=\,
\psi_a^{-1}\,(\,e_{1}\psi_b\,)\,\,\,\,\,\mbox{and}\,\,\,\,\,\lambda=
1 +\,\psi_a^{-1}\,(\,e_{1}\psi_b\,)\,\,.\] Restricting ourselves
to quaternionic sub-algebras containing $e_{1}$, i.e.
\[e_{1}e_{2}e_{3}\,\,\,,\,\,\,\,\,
e_{1}e_{4}e_{5}\,\,\,\,\,\mbox{and}\,\,\,\,\,
e_{1}e_{7}e_{6}\,\,,
\] we immediately find the following solutions
\[
\begin{array}{lclcl}
\left\{\, \psi_{a}\, ,\, \psi_{b}\,;\,\lambda
\,\right\}_{1,2}&=& \left\{\, e_{2}\, ,\, e_{3}\,;\,0
\,\right\}_{1} &\,\,\mbox{and}\,\,&
\left\{\, e_{2}\, ,\, -\, e_{3}\,;\,2 \,\right\}_{2}\,\,,\\
& & \left\{\, e_{4}\, ,\, e_{5}\,;\,0 \,\right\}_{1} & &
\left\{\, e_{4}\, ,\, -\, e_{5}\,;\,2 \,\right\}_{2}\,\,,\\
& & \left\{\, e_{7}\, ,\, e_{6}\,;\,0 \,\right\}_{1} & &
\left\{\, e_{7}\, ,\, -\, e_{6}\,;\,2 \,\right\}_{2}\,\,.
\end{array}
\]
To simplify our discussion, in this example we choose  the
quaternionic sub-algebra $e_{1} e_{2} e_{3}$. It is rather
surprising that, notwithstanding the right position of the
octonionic eigenvalue, we find, as for the case of the {\sc QlEP},
new solutions characterized by {\em not} real eigenvalues:
\[
\begin{array}{lclcl}
\left\{\, \psi_{a}\, ,\, \psi_{b}\,;\,\lambda
\,\right\}_{1,10}&=& \left\{\, e_{2}\, ,\, e_{3}\,;\,0
\,\right\}_{1} &\,\,\mbox{and}\,\,&
\left\{\, e_{2}\, ,\, -\, e_{3}\,;\,2 \,\right\}_{2}\,\,,\\
& & \left\{\, e_{2}\, ,\, e_{4}\,;\,1-e_{7} \,\right\}_{3} & &
\left\{\, e_{2}\, ,\, -\, e_{4}\,;\, 1 + e_{7} \,\right\}_{4}\,\,,\\
& & \left\{\, e_{2}\, ,\, e_{5}\,;\,1+e_{6} \,\right\}_{5} & &
\left\{\,
e_{2}\, ,\, -\, e_{5}\,;\,1-e_{6} \,\right\}_{6}\,\,,\\
& & \left\{\, e_{2}\, ,\, e_{6}\,;\,1-e_{5} \,\right\}_{7} & &
\left\{\, e_{2}\, ,\, -\, e_{6}\,;\,1+e_{5} \,\right\}_{8}\,\,,\\
& & \left\{\, e_{2}\, ,\, e_{7}\,;\,1+e_{4} \,\right\}_{9} & &
\left\{\, e_{2}\, ,\, -\, e_{7}\,;\,1-e_{4}
\,\right\}_{10}\,\,.
\end{array}
\]
This result is well known in literature. For a deeper discussion
of similar problems encountered in studying octonionic eigenvalue
equations  for Hermitian matrices and for mathematical techniques
to find the new {\em not\,} real eigenvalues, the best references
are\,\cite{DM98a,DM98b,DM99,DJM00,DMO02}. In this paper, we are
interested in presenting a conversion method, in the main quite
practical, to find real eigenvalues of octonionic operators. The
matrix representations for these operators\,\cite{DK97,DD99} could
be very useful  in view of possible applications in quantum
mechanics. Before to elaborate the core of our article, let us
briefly explain where the proof given for the {\sc CEP} and for
the {\sc QrEP} is lacking in validity. Due the non-associativity
of the octonionic algebra
\[ \Psi^{\dag} (M \Psi) \neq (M \Psi)^{\dag}\Psi \,[= (\Psi^{\dag}
M)\Psi] \,\, .\] Consequently, the position of the octonionic
eigenvalue in the {\sc OrEP} is {\em not} sufficient to guarantee
the real nature of Hermitian matrix eigenvalues.

The results of this preliminary study are instructive for several
reasons. In the first place, they show that using a quaternionic
algebra a great simplification can be obtained by making an
initial assumption concerning the position of the quaternionic
eigenvalue. Secondly, the possibility to apply a similarity
transformation ($u\lambda \bar{u}=z$) opens the door to translate
the {\sc QrEP} in its complex or real counterpart. This method is
frequently employed to circumvent the difficulties attendant the
non-commutativity of quaternions\,\cite{DDu99}. What we aim to prove
in this paper is a stronger result. The existence of conversion
rules for octonionic operators allow to give a practical method to
solve the coupled octonionic eigenvalue problem ({\sc cOEP}) which
reduces to the {\sc QrEP} in the quaternionic limit~\,\cite{DS00}.

\section{Octonionic eigenvalue problem}

\subsection{Real matrix conversion}
\label{rmt}

The non-associativity of octonions seems to suggest the
impossibility to obtain a real matrix representations (with the
standard matrix multiplication rules) for octonionic
operators\,\cite{DD99}. Nevertheless, the use of left/right barred
octonionic operators allow to reproduce the $GL(8,\mathbb{R})$
group\,\cite{DK97}. For the convenience of the reader and to make
our exposition as self-contained as possible, we repeat part of
the  material exposed in ref.\,\cite{DK97}.

It's well known that octonions are a non commutative algebra so we
must distinguish between left and right actions of the octonionic
imaginary units $e_m$, by introducing operators $L_{o_1}$ and
$R_{o_2}$ whose action on octonionic functions of a real variable,
$\psi$, $\psi: \mathbb{R} \to \mathbb{O}$, gives
\begin{equation}
L_{o_1} \psi = o_1 \, \psi~~~\mbox{and}~~~ R_{o_2} \psi =\psi \,
o_2~,
\end{equation}
But, octonions are also a non associative algebra that means
\[
o_1 (\psi o_2) \neq (o_1 \psi) o_2
\]
so $L_{o_1}\,R_{o_2}\, \psi$ must be calculated, necessarily, in
the order the operations appear, that is,
\[
L_{o_1}\,R_{o_2}\, \psi = L_{o_1} ( R_{o_2}\, \psi )= o_1 (\psi
o_2)
\]
and
\[
R_{o_2} \, L_{o_1} \, \psi = R_{o_2} ( L_{o_1} \, \psi) = ( o_1
\psi) o_2
\]
Naturally the same holds for
\[
L_{o_1} L_{o_2} \psi = L_{o_1} (L_{o_2} \psi) = o_1 (o_2 \psi)
\]
and
\[
R_{o_1} R_{o_2} \psi = R_{o_1} (R_{o_2} \psi) = R_{o_1} (\psi o_2)
= (\psi o_2) o_1
\]
In order to write the more general octonionic operator it's enough
to describe all the possible actions of the imaginary units $e_m$.
So, computing the operators just described for the imaginary units
we find
\begin{equation}
\label{106}
\begin{array}{cccl}
1, L_m, R_m &  & & 15 ~\mbox{elements}\\
L_m R_m = R_m L_m & & &  ~\,7~ \mbox{elements}\\
L_m R_n~~m \neq n & & & 42~ \mbox{elements}\\
R_n L_m~~m \neq n & & & 42~ \mbox{elements}
\end{array}
\end{equation}
for $m,n=1,\cdots,7$, which totalizes 106 operators. However, it
is possible to prove that $L_m R_n$, can be expressed by a
suitable combination of $R_n L_m$ operators, reducing to 64 the previous 106 elements\,\cite{DK97}. For example, we have
\[
L_m R_n + L_n R_m = R_n L_m + R_m L_n\,\,.
\]
As explicitly shown in Appendix A of ref.\,\cite{DK97}, it is possible to prove the linear independence of these 64 elements which  represent the most general octonionic operator
\begin{equation}
\label{geno} o_{_0} + \sum_{m=1}^{7} R_{_m} \,L_{o_m}
\end{equation}
with $o_0, \cdots , o_7 \in \mathbb{O}$. This  shows the correspondence between our
generalized octonions, equation (\ref{geno}), and $GL(8,
\mathbb{R})$.

In paper \,\cite{DD99}, the authors give representations of
octonions and other nonassociative algebras by special
matrices, which are endowed with special multiplication rules. The
introduction of left/right octonionic operators allow us to
establish the isomorphism between themselves and $GL(8,
\mathbb{R})$ with the standard multiplication rules. In order to
explain the idea of conversion, let us look explicitly at the
action of the operators $R_1$ and $L_2$ on a generic octonionic
function $\psi: \mathbb{R} \to \mathbb{O}$:\\

\hspace*{-8mm}
$
\psi(x) = \psi_0(x) + e_1 \psi_1(x) + e_2 \psi_2(x)+ e_3
\psi_3(x)+ e_4 \psi_4(x)+ e_5 \psi_5(x)+ e_6 \psi_6(x)+ e_7
\psi_7(x)
$\\

\noindent
with $\psi_{0, \cdots ,7}:\mathbb{R} \to \mathbb{R}$. In order to
simplify our notation we omit $x$ variable. So, let us calculate
$R_1 \psi$ that gives
\[
R_1 \psi = \psi e_1 = e_1 \psi_0 - \psi_1 - e_3 \psi_2 + e_2
\psi_3 - e_5 \psi_4 + e_4 \psi_5 + e_7 \psi_6 - e_6 \psi_7
\]
and now, we calculate
\[
L_2 \psi = e_2\psi = e_2 \psi_0 - e_3 \psi_1 - \psi_2 + e_1 \psi_3
+ e_6 \psi_4 + e_7 \psi_5 - e_4 \psi_6 - e_5 \psi_7
\]
If we represent the octonionic function $\psi$ by a real column
vector $8 \times 1$
\begin{equation}
\psi~~ \leftrightarrow~~\Psi = \left( \begin{array}{c}
                           \psi_0 \\
                           \psi_1 \\
                           \psi_2 \\
                           \psi_3 \\
                           \psi_4 \\
                           \psi_5\\
                           \psi_6 \\
                           \psi_7
                                 \end{array}   \right)
\end{equation}
we can rewrite $\psi e_1$ and $e_2 \psi$ respectively as
\[
\left(
\begin{array}{rrrrrrrr}
0 & -1 & 0 & 0 & 0 & 0 & 0 & 0\\
1 &  0 & 0 & 0 & 0 & 0 & 0 & 0\\
0 &  0 & 0 & 1 & 0 & 0 & 0 & 0\\
0 &  0 & -1 & 0 & 0 & 0 & 0 & 0\\
0 &  0 & 0 & 0 & 0 & 1 & 0 & 0\\
0 &  0 & 0 & 0 & -1 & 0 & 0 & 0\\
1 &  0 & 0 & 0 & 0 & 0 & 0 & -1\\
1 &  0 & 0 & 0 & 0 & 0 & 1 & 0
\end{array}  \right)\left( \begin{array}{c}
                           \psi_0 \\
                           \psi_1 \\
                           \psi_2 \\
                           \psi_3 \\
                           \psi_4 \\
                           \psi_5\\
                           \psi_6 \\
                           \psi_7
                                 \end{array}   \right) = \left( \begin{array}{r}
                           -\psi_1 \\
                           \psi_0 \\
                           \psi_3 \\
                           -\psi_2 \\
                           \psi_5 \\
                           -\psi_4\\
                           -\psi_7 \\
                           \psi_6
                                 \end{array}   \right)
\]
and
\[
\left(
\begin{array}{rrrrrrrr}
0 &  0 & -1 & 0 & 0 & 0 & 0 & 0\\
0 &  0 & 0 & 1 & 0 & 0 & 0 & 0\\
1 &  0 & 0 & 0 & 0 & 0 & 0 & 0\\
0 &  -1 & 0 & 0 & 0 & 0 & 0 & 0\\
0 &  0 & 0 & 0 & 0 & 0 & -1 & 0\\
0 &  0 & 0 & 0 & 0 & 0 & 0 & -1\\
0 &  0 & 0 & 0 & 1 & 0 & 0 & 0\\
0 &  0 & 0 & 0 & 0 & 1 & 0 & 0
\end{array}  \right)\left( \begin{array}{c}
                           \psi_0 \\
                           \psi_1 \\
                           \psi_2 \\
                           \psi_3 \\
                           \psi_4 \\
                           \psi_5\\
                           \psi_6 \\
                           \psi_7
                                 \end{array}   \right) = \left(
\begin{array}{r}
                           -\psi_2 \\
                           \psi_3 \\
                           \psi_0 \\
                           -\psi_1 \\
                           -\psi_6 \\
                           -\psi_7\\
                           \psi_4 \\
                           \psi_5
                                 \end{array}   \right)
\]
Following this procedure, we can construct the complete set of
conversion rules for the imaginary units operators $L_m$ and
$R_m$. Observe that if we multiply the matrices that represent
$L_1$ and $L_2$ we will find a new matrix that is different from
$L_3$, that is, $L_1 L_2 \neq L_3$ while $e_1 e_2 = e_3$. This
bluff is soon explained. In deducing our conversion rules, we
understand octonions as operators, and so they must be applied to
a certain octonionic function, $\psi$. If we have the octonionic
relation
\[
(e_1 e_2)\psi =  e_3 \psi
\]
the matrix counterpart will be
\[
L_3 \psi
\]
since the matrix counterparts are defined by their action upon a
function and not upon another operator. On the other hand,
\[
e_1 ( e_2 \psi) \neq e_3 \psi
\]
will be translated by
\[
L_1 L_2 \psi = L_1 (L_2 \psi) \neq L_3 \psi
\]
We have to differentiate between two kinds of multiplication, one,
for octonions, denoted by a middle dot, "$\cdot$", and the other,
for octonionic operators, denoted by the usual multiplication sign
"$\times$". At the level of octonions, one has
\[
e_1 \cdot e_2 = e_3
\]
but at the level of octonionic operators
\[
L_1 \times L_2 \neq L_3
\]
but
\[
L_1 \times L_2 = L_3 +  R_2 L_1 -  L_1 R_2
\]
Observe the matricial representations of our operators enable us
to reproduce the octonion nonassociativity by the matrix algebra.
Consider, for example,
\begin{eqnarray}
R_1 L_3 \psi = R_1 ( L_3 \psi ) = (e_3 \psi) e_1 =~~~~~~~~~~ \nonumber \\
e_2 \psi_0 - e_3 \psi_1 + \psi_2 - e_1 \psi_3 - e_6 \psi_4 - e_7
\psi_5 + e_4 \psi_6 + e_5 \psi_7
\end{eqnarray}
which gives, in matricial representation
\[
\left(
\begin{array}{rrrrrrrr}
0 &  0 & 1 & 0 & 0 & 0 & 0 & 0\\
0 &  0 & 0 & -1 & 0 & 0 & 0 & 0\\
1 &  0 & 0 & 0 & 0 & 0 & 0 & 0\\
0 &  -1 & 0 & 0 & 0 & 0 & 0 & 0\\
0 &  0 & 0 & 0 & 0 & 0 & 1 & 0\\
0 &  0 & 0 & 0 & 0 & 0 & 0 & 1\\
0 &  0 & 0 & 0 & -1 & 0 & 0 & 0\\
0 &  0 & 0 & 0 & 0 & -1 & 0 & 0
\end{array}  \right)\left( \begin{array}{c}
                           \psi_0 \\
                           \psi_1 \\
                           \psi_2 \\
                           \psi_3 \\
                           \psi_4 \\
                           \psi_5\\
                           \psi_6 \\
                           \psi_7
                                 \end{array}   \right) = \left(
\begin{array}{r}
                           \psi_2 \\
                           -\psi_3 \\
                           \psi_0 \\
                           -\psi_1 \\
                           \psi_6 \\
                           \psi_7\\
                           -\psi_4 \\
                           -\psi_5
                                 \end{array}   \right)
\]
whereas
\begin{eqnarray}
L_3 R_1 \psi = L_3 ( R_1 \psi ) = e_3 (\psi e_1) =~~~~~~~~~~ \nonumber\\
e_2 \psi_0 - e_3 \psi_1 + \psi_2 - e_1 \psi_3 + e_6 \psi_4 + e_7
\psi_5 - e_4 \psi_6 - e_5 \psi_7
\end{eqnarray}
which gives, in matricial representation
\[
\left(
\begin{array}{rrrrrrrr}
0 &  0 & 1 & 0 & 0 & 0 & 0 & 0\\
0 &  0 & 0 & -1 & 0 & 0 & 0 & 0\\
1 &  0 & 0 & 0 & 0 & 0 & 0 & 0\\
0 &  -1 & 0 & 0 & 0 & 0 & 0 & 0\\
0 &  0 & 0 & 0 & 0 & 0 & -1 & 0\\
0 &  0 & 0 & 0 & 0 & 0 & 0 & -1\\
0 &  0 & 0 & 0 & 1 & 0 & 0 & 0\\
0 &  0 & 0 & 0 & 0 & 1 & 0 & 0
\end{array}  \right)\left( \begin{array}{c}
                           \psi_0 \\
                           \psi_1 \\
                           \psi_2 \\
                           \psi_3 \\
                           \psi_4 \\
                           \psi_5\\
                           \psi_6 \\
                           \psi_7
                                 \end{array}   \right) = \left(
\begin{array}{r}
                           \psi_2\\
                           -\psi_3\\
                           \psi_0\\
                           -\psi_1\\
                           -\psi_6\\
                           -\psi_7\\
                           \psi_4\\
                           \psi_5
                                 \end{array}   \right)
\]
From the 106 elements given in (\ref{106}), that we can
rewrite in the matricial form, we can extract two different basis
bases for $GL(8,\mathbb{R})$ those are
\[
1, L_m, R_n, R_n L_m
\]
or
\[
1, L_m, R_n, L_m R_n
\]
We now remark some difficulties deriving from octonion
nonassociativity. When we translate from generalized octonions to
$8 \times 8$ real matrices there is no problem. For example, in
the octonionic object
\[
e_4 \{ [(e_6 \psi )e_1]e_5 \}
\]
we quickly recognize the operators
\[
L_4 R_5~~~\mbox{and}~~~R_1 L_6~.
\]
Thus, rewriting the previous object we have
\[
L_4 R_5 \, R_1 L_6 \psi
\]
In going from $8 \times 8$ real matrices to
octonions, we should be careful in ordering. For example,
\[
AB\,\psi
\]
can be understood as
\[
(AB)\,\psi~~~\mbox{or}~~~A \,(B \,\psi)~.
\]
For example, by choosing $A=L_{1}$ and $B=L_2$, we have two possible different translations
\[(L_1L_2)\,\psi\,\,\to\,\,(e_1e_2)\,\psi=e_3\,\psi~~~\mbox{or}~~~ L_1(L_2\psi)\,\,\to\,\,e_1\,(e_2\psi)\,\,.\]
Which is the right equation? The second translation
is the right one, observe that  $L_1L_2\neq L_3$. So, when we find multiple
multiplications, to translate correctly in the octonionic formalism, we have to use the following ordering rule
\[
ABC \dots Z \psi = A ( B ( \dots (Z \psi)))\,\,.
\]

\subsection{Coupled eigenvalue problem}
\label{title-cep}

Due to the non commutativity of quaternions, they have represented
a challenge when mathematicians tried to extend the well-known
complex eigenvalue problem to the quat\-ernionic field. The
eigenvalue problem for $\mathbb{H}$-linear quaternionic operators,
which means that only the left action of quaternionic imaginary
units are present, and an extension to $\mathbb{C}$ and
$\mathbb{R}$-linear quaternionic operator that are, respectively,
operators which contain, besides the left action of all units, the
right action of only one imaginary unit and the right action of
all quaternionic imaginary units, has been recently discussed and
can be found in \,\cite{DS00,DSS02}. The quaternionic eigenvalue
problem allowed the first steps in the theory of quaternionic
differential equations \,\cite{DD01} and, the latter, made possible
the study of the Schr\"odinger equation in the presence of a
quaternionic potential in view of analyze the quaternionic
tunnelling effect \,\cite{DD05}. It's worth pointing out that the
eigenvalue problem for $\mathbb{R}$-linear quaternionic operators
provides a remarkable version of the eigenvalue problem dividing
it in two coupled equations. For the octonionic eigenvalue problem
we resort to this same method that gives out a coupled eigenvalue
problem.

Consider the octonionic real eigenvalue problem
\[
\mathcal{O} \psi = \psi r\,,~~~~~r
\in \mathbb{R}
\]
which matricial conversion is given by
\begin{equation}
\label{oep} M_o \Psi_{8 \times 1} = r \, \Psi_{8 \times 1}~,
\end{equation}
where $M_o$ is a real square matrix of order 8. Note that the
eigenvalue must be real and this condition is too restrictive. The
matrix $M_o \in M_8[\mathbb{R}]$ will probably have complex
eigenvalues and, consequently, complex eigenvectors but when this
occur the conversion backward to the octonionic formalism is not
possible. This is exactly the same problem when studying the
eigenvalue problem for real linear quaternionic operators. For a
deeper discussion of this subject we refer the reader to
\,\cite{DSS02}. We will use the same technique to study the
octonionic eigenvalue problem but here this is done in a much more
roughly way. In order to differentiate between octonionic and
complex formalisms, after the conversion from octonion to
matrices, the complex unit will be denoted by $i$. So, accepting
that $M_o$ has complex eigenvalues we can rewrite the eigenvalue
problem (\ref{oep}) as
\begin{equation}
\label{oepmod}
 M_o \Psi = z \, \Psi~,~~z \in \mathbb{C}~.
\end{equation}
Introducing $\Psi = \xi + i \eta$, where $\xi$ and $\eta$ are real
column vectors $8 \times 1$ and $z = a + i b$, with $a, b \in
\mathbb{R}$ in (\ref{oepmod}) we have
\[
M_o ( \xi + i \eta ) = (a + i b)( \xi + i \eta )
\]
Separating the real from the imaginary part we have a coupled
equation given by
\begin{equation}
\label{coep}
\begin{array}{l}
M_o \xi = a \xi - b \eta\\
M_o \eta = a \eta + b \xi
\end{array}
\end{equation}
Now, all the elements present in (\ref{coep})
are real which
allows the conversion to the octonionic formalism. Let see a one
dimensional example. Consider the problem
\[
e_4 \psi = \psi \lambda~.
\]
We want to find $\xi, \eta$ and $\lambda \in \mathbb{C}$ that
satisfies (\ref{coep}). The first step is to translate the problem
in the matricial form which is given by
\begin{equation}
\label{mcoep}
\left( \begin{array}{rrrrrrrr}
0 & 0 & 0 & 0 & -1 & 0 & 0 & 0\\
0 & 0 & 0 & 0 & 0 & 1 & 0 & 0\\
0 & 0 & 0 & 0 & 0 & 0 & 1 & 0\\
0 & 0 & 0 & 0 & 0 & 0 & 0 & 1\\
1 & 0 & 0 & 0 & 0 & 0 & 0 & 0\\
0 & -1 & 0 & 0 & 0 & 0 & 0 & 0\\
0 & 0 & -1 & 0 & 0 & 0 & 0 & 0\\
0 & 0 & 0 & -1 & 0 & 0 & 0 & 0
\end{array} \right)
\left( \begin{array}{r}
                   \psi_0\\
                   \psi_1\\
                   \psi_2\\
                   \psi_3\\
                   \psi_4\\
                   \psi_5\\
                   \psi_6\\
                   \psi_7
                 \end{array} \right) = z \left( \begin{array}{r}
                   \psi_0\\
                   \psi_1\\
                   \psi_2\\
                   \psi_3\\
                   \psi_4\\
                   \psi_5\\
                   \psi_6\\
                   \psi_7
                 \end{array} \right)~.
\end{equation}
The eigenvalues and eigenvectors of the problem (\ref{mcoep}) are, respectively,\\

\noindent eigenvalues:\vspace*{2mm}

$
\begin{array}{lcccccccc}
 & ~-i~~ & ~~~-i~~~ & ~~-i~~~ & ~~~-i~~~ & ~~~~~~i~~~~ & ~~~~~~i~~~~ & ~~~~~~i~~~~ & ~~~~~~i~~~~\\
\end{array}
$

\noindent eigenvectors:\vspace*{2mm}

$
\begin{array}{lcccccccc}
&  \left( \begin{array}{r}
                   0\\
                   0\\
                   0\\
                   i\\
                   0\\
                   0\\
                   0\\
                   1
                 \end{array} \right) & \left( \begin{array}{r}
                   0\\
                   0\\
                   i\\
                   0\\
                   0\\
                   0\\
                   1\\
                   0
                 \end{array} \right)& \left( \begin{array}{r}
                   0\\
                   i\\
                   0\\
                   0\\
                   0\\
                   1\\
                   0\\
                   0
                 \end{array} \right) & \left( \begin{array}{r}
                   -i\\
                   ~0\\
                   ~0\\
                   ~0\\
                   ~1\\
                   ~0\\
                   ~0 \\
                   ~0
                 \end{array} \right) & \left( \begin{array}{r}
                   ~0\\
                   ~0\\
                   ~0\\
                   -i\\
                   ~0\\
                   ~0\\
                   ~0\\
                   ~1
                \end{array} \right) & \left( \begin{array}{r}
                   ~0\\
                   ~0\\
                   -i\\
                   ~0\\
                   ~0\\
                   ~0\\
                   ~1\\
                   ~0
                 \end{array} \right) & \left( \begin{array}{r}
                   ~0\\
                   -i\\
                   ~0\\
                   ~0\\
                   ~0\\
                   ~1\\
                   ~0\\
                   ~0
                 \end{array} \right) & \left( \begin{array}{r}
                    i\\
                    0\\
                    0\\
                    0\\
                    1\\
                    0\\
                    0\\
                    0
                 \end{array} \right)
                 \end{array}
$\\[2mm]

\noindent Consider $\lambda = -i$ and the respective eigenvector $\Psi = (
0~, 0~, 0~, i~, 0~, 0~, 0~, 1)^t$. We have $a = 0$ and $b = -1$
and separating the real from the imaginary part of the
eigenvector,
\[
\left( \begin{array}{c}
                   0\\
                   0\\
                   0\\
                   i\\
                   0\\
                   0\\
                   0\\
                   1
\end{array} \right) = \left( \begin{array}{c}
                   0\\
                   0\\
                   0\\
                   0\\
                   0\\
                   0\\
                   0\\
                   1
                \end{array} \right) + i ~\left( \begin{array}{c}
                   0\\
                   0\\
                   0\\
                   1\\
                   0\\
                   0\\
                   0\\
                   0
                  \end{array} \right)
\]
we obtain
\[
\xi = \left( \begin{array}{c}
                   0\\
                   0\\
                   0\\
                   0\\
                   0\\
                   0\\
                   0\\
                   1
                   \end{array} \right)~~~\mbox{and}~~~\eta = \left( \begin{array}{c}
                   0\\
                   0\\
                   0\\
                   1\\
                   0\\
                   0\\
                   0\\
                   0
                   \end{array} \right)
\]
Note that all elements $a, b, \xi$ and $\eta$ are real so we can
transform each of them in octonionic numbers. This procedure gives
\[
\xi = e_7~~~\mbox{and}~~~\eta = e_3
\]
and the coupled equations, already in the octonionic formalism,
became
\begin{eqnarray*}
e_4 (e_7) = ~0 \, e_3 + 1 \, e_3 = ~e_3\\
e_4 (e_3) = - \,e_7 + 0 \, e_3 = - e_7
\end{eqnarray*}
Following the same procedure, consider the $2 \times 2$ octonionic
matrix
\[
\left( \begin{array}{rr}
  1 & e_4\\
  0 & e_5
\end{array} \right)~.
\]
Its conversion lead us to a $16 \times 16$ real matrix which
eigenvalues are $i$, $-i$ and $1$. The algebraic multiplicity of
$\lambda$ is equal to its geometric multiplicity. In this case $i$
and $-i$ have algebraic multiplicity equal 4 and $1$ has algebraic
multiplicity equal 8. By taking the eigenvalue $-i$ we have $a=0$
and $b = -1$ and the correspondent real vectors ($\xi$ and $\eta$)
are given by $\xi = ( 0~, 0~, 0~, -1~, 0~, 0~, 1~, 0~, 0~, 0~,0~,
0~, 0~,0~, 0~, 2)^t$ and $\eta = ( 0~, 0~, 0~, -1~, 0~, 0~, 1~,
0~, 0~, 0~,2~, 0~, 0~,0~, 0~, 0)^t$ which, in octonionic formalism
are written as
\[
\xi = \left(
\begin{array}{c}
- e_3 + e_6\\
2 e_7
\end{array} \right)~~~\mbox{and}~~~\eta = \left(
\begin{array}{c}
e_3 + e_6\\
2 e_2
\end{array} \right)
\]
and the coupled equation became
\begin{eqnarray*}
\left(
\begin{array}{rr}
  1 & e_4\\
  0 & e_5
\end{array} \right)
\left(
\begin{array}{c}
- e_3 + e_6\\
2 e_7
\end{array} \right) = 0 \left( \begin{array}{c}
- e_3 + e_6\\
2 e_7
\end{array} \right) - (-1)
\left(
\begin{array}{c}
e_3 + e_6\\
2 e_2
\end{array} \right)\\
\\
\left(
\begin{array}{rr}
  1 & e_4\\
  0 & e_5
\end{array} \right)
\left(
\begin{array}{c}
e_3 + e_6\\
2 e_2
\end{array} \right) = -
\left(
\begin{array}{c}
- e_3 + e_6\\
2 e_7
\end{array} \right) + 0 \left(
\begin{array}{c}
e_3 + e_6\\
2 e_2
\end{array} \right)
\end{eqnarray*}

It's interesting to note that when the eigenvalue problem is
considered for a $\mathbb{H}$-linear quaternionic operator, we
have
\begin{equation}
\label{qe}
 \mathcal{Q} \psi = \psi \lambda = \psi a + \psi ib~.
\end{equation}
Multiplying the expression above by $i$ we obtain
\[
(\mathcal{Q} \psi) i = (\psi a) i + (\psi \, i b) i~,
\]
but quaternions are associative which means that
\[
\mathcal{Q} (\psi \, i) = (\psi i) a - \psi b~.
\]
The eigenvectors $\psi \, i$ and $\psi$ play the role of $\xi$ and
$\eta$, respectively, in the coupled equations. Thus,
\begin{equation}
\label{qce}
\begin{array}{l}
\mathcal{Q} \xi = \mathcal{Q} (\psi i) = (\psi i) a - \psi b~ =
\xi a - \eta b\\
\mathcal{Q} \eta = ~\mathcal{Q} \psi~~ = ~\psi a + (\psi i)b =
\eta a + \xi b
\end{array}
\end{equation}
which means that it's possible to associate to any pair of
functions $\xi$ and $\eta$ that solves the system (\ref{qce}) a
corresponding eigenvector of (\ref{qe}).

\subsection{Complexified eigenvalue problem}

An alternative way to solve octonionic eigenvalue problems is to consider complexified octonions, $\mathbb{C}(1,i) \times \mathbb{O}$, which allow to immediately
translate Eq. (\ref{oepmod}) as follows
\[
\mathcal{O} \Phi = \Phi\,z~,
\]
where $\mathcal{O}$ is a octonionic matrix, $\Psi \in \mathbb{C}(1,i) \times \mathbb{O}$ and $z \in \mathbb{C}(1,i)$. Introducing
$\Phi = \phi_1 + i \phi_2$ and $z = a + i b$ we have
\begin{equation}
\label{comp-prob}
\mathcal{O} ( \phi_1 + i \phi_2 ) =  ( \phi_1 + i \phi_2 ) (a + i b)~,\,\,\,\,\, \mathcal{O},\phi_1,\phi_2 \in \mathbb{O}~,\,\,\,\,\, a, b \in \mathbb{R}\,\,.
\end{equation}
Algebraic manipulations lead us to the solution of the problem. For example, consider the same problem presente in Section \ref{title-cep}, which is
\begin{equation}
\label{ex-compl}
e_4 ( \phi_1 + i \phi_2 )  = ( \phi_1 + i \phi_2 ) (a + i b)~.
\end{equation}
Now, multiplying (\ref{ex-compl}) by $e_4$ from the left we find
\[
e_4 \left[ e_4 ( \phi_1 + i \phi_2 ) \right] =  e_4 \left[ ( \phi_1 + i \phi_2 ) (a + i b) \right]~,
\]
which gives
\[
- ( \phi_1 + i \phi_2 ) = \left[ e_4 ( \phi_1 + i \phi_2 ) \right] (a + ib) = (a + ib)^2 ( \phi_1 + i \phi_2 )~,
\]
since $e_4$ commutes with $a+ib$. Equation above gives the very simple system
\[
a^2 - b^2 = -1~~~\mbox{and}~~~2 a b = 0
\]
which solution is $a=0$ and $b = \pm 1$. So, the eigenvalue is $\pm i$. Now, reintroducing $-i$ in the eigenvalue equation we find
the eigenvectors. Then, for $z = -i$ we have
\[
e_4 ( \phi_1 + i \phi_2 ) = ( \phi_1 + i \phi_2 ) (-i) = - i \phi_1 + \phi_2 ~~~\Rightarrow~~~ e_4 \phi_1 =  \phi_2~, e_4 \phi_2 = -\phi_1
\]
Remembering that the octonionic imaginary units obey the relation
\begin{equation}
123,~145,~176,~246,~257,~347~~~\mbox{and}~~~365\,,
\end{equation}
it's easy to see that $\phi_1$ can assume the values $e_4, e_5, e_6$ and $e_7$ and $\phi_2$ can assume, respectively,
the values $-1, e_1, e_2$ and $e_3$. Thus
\begin{eqnarray*}
e_4 (e_4 - i) &=& (e_4 - i)(-i)~,\\
e_4 (e_5 + e_1) &=& (e_5 + e_1)(-i)~,\\
e_4 (e_6 + e_2) &=& (e_6 + e_2)(-i)~,\\
e_4 (e_7 + e_3) &=& (e_7 + e_3)(-i)~.\\
\end{eqnarray*}

Let us consider the complexified octonion formalism to solve another example from section \ref{title-cep}. Consider
\[
\left(
\begin{array}{rr}
  1 & e_4\\
  0 & e_5
\end{array} \right)~.
\]
We want to find the eigenvalues $\lambda = a + ib~, a, b \in \mathbb{R}$ and the respective eigenvectors which have
the following form
\[
\begin{pmatrix}
\phi_1\\
\phi_2
\end{pmatrix} + i \begin{pmatrix}
\psi_1\\
\psi_2
\end{pmatrix}~,~~\phi_{1,2}, \psi_{1,2} \in \mathbb{O}~.
\]
The eigenvalue problem is
\[
\left(
\begin{array}{rr}
  1 & e_4\\
  0 & e_5
\end{array} \right)  \begin{pmatrix}
  \phi_1 + i \psi_1\\
  \phi_2 + i \psi_2
\end{pmatrix} = \begin{pmatrix}
  \phi_1 + i \psi_1\\
  \phi_2 + i \psi_2
\end{pmatrix} (a + i b)~.
\]
Performing the matrix multiplication, we have
\begin{eqnarray}
\label{eq1}
e_4 (\phi_2 + i \psi_2) & = & (\phi_1 + i \psi_1) (a - 1 + ib)~,\\
\label{eq2}
e_5 (\phi_2 + i \psi_2) & = & (\phi_2 + i \psi_2) (a + ib)
\end{eqnarray}
The proceedure to solve eq.(\ref{eq2}) is almost the same as the previously just done. So, repeating it
and fixing the eigenvalue $-i$ we find
\[
\phi_2 + i \psi_2 \in \{ 1 + i e_5~,~e_1 + i e_4~,~e_3 + i e_6~,~e_7 + i e_2 \}
\]
In order to find $\phi_1$ and $\psi_1$ we have to fix an eigenvalue on the set above which introduced in
eq.(\ref{eq1}) gives the result. To find all the possibilities we have to do this for all elements of the set.
To illustrate this, we choose $1 + i e_5$. Remenber that we are working with $\lambda=-i$, so
\[
e_4 (1 + i e_5)  =  (\phi_1 + i \psi_1) (-1 - i) \Rightarrow e_4 + i e_1 = - \phi_1 - i \phi_1 - i \psi _1 + \psi_1
\]
Solving the system of equations we find $\phi_1 = - (e_1 + e_4)/2$ and $\psi_1 = (e_4 - e_1)/2$ and we have\\

\hspace*{-3mm}
$
\left(
\begin{array}{rr}
  1 & e_4\\
  0 & e_5
\end{array} \right) \left[ \left( \begin{array}{c}
  - e_1 -  e_4\\
        2
\end{array} \right) + i \left( \begin{array}{c}
e_4 - e_1\\
2e_5
\end{array} \right) \right]
= \left[ \left( \begin{array}{c}
  - e_1 -  e_4\\
        2
\end{array} \right) + i \left( \begin{array}{c}
e_4 - e_1\\
2e_5
\end{array} \right) \right](-i)~.
$\\

\noindent Repeating the procedure for the remaining three cases, always for $\lambda = -i$ we have\\

\hspace*{-3mm}
$
\left(
\begin{array}{rr}
  1 & e_4\\
  0 & e_5
\end{array} \right) \left[ \left( \begin{array}{c}
  1 + e_5\\
  2e_1
\end{array} \right) + i \left( \begin{array}{c}
1 - e_5\\
2 e_4
\end{array} \right) \right]
= \left[ \left( \begin{array}{c}
  1 + e_5\\
  2e_1
\end{array} \right) + i \left( \begin{array}{c}
1 - e_5\\
2 e_4
\end{array} \right) \right](-i)~,
$

\vspace*{5mm}

\hspace*{-3mm}
$
\left(
\begin{array}{rr}
  1 & e_4\\
  0 & e_5
\end{array} \right) \left[ \left( \begin{array}{c}
  e_2 - e_7\\
  -2e_3
\end{array} \right) + i \left( \begin{array}{c}
e_2 + e_7\\
-2 e_6
\end{array} \right) \right]
= \left[  \left( \begin{array}{c}
  e_2 - e_7\\
  -2e_3
\end{array} \right) + i \left( \begin{array}{c}
e_2 + e_7\\
-2 e_6
\end{array} \right) \right] (-i)~,
$

\vspace*{5mm}

\hspace*{-3mm}
$
\left(
\begin{array}{rr}
  1 & e_4\\
  0 & e_5
\end{array} \right) \left[ \left(\begin{array}{c}
  e_6 - e_3\\
  2e_7
\end{array} \right) + i \left( \begin{array}{c}
e_3 + e_6\\
2 e_2
\end{array} \right) \right]
= \left[ \left(\begin{array}{c}
  e_6 - e_3\\
  2e_7
\end{array} \right) + i \left( \begin{array}{c}
e_3 + e_6\\
2 e_2
\end{array} \right) \right] (-i)~.
$

\vspace*{5mm}
\noindent Observe that what is missing when comparing the procedure, for a matrix of order 2,  just presented with the previous case is that, in addition, from eq.(\ref{eq2}) we have
\[
\phi_2 + i \psi_2 = 0~~~\Rightarrow~~~\phi_2 = \psi_2 = 0~.
\]
Introducing $\phi_2 + i \psi_2 = 0$ in eq.(\ref{eq1}) we obtain
\[
(\phi_1 + i \psi_1) (a - 1 + ib) = 0~~~\Rightarrow~~~a=1~, b = 0~.
\]
So, another eigenvalue of the octonionic problem discussed is $\lambda = 1$ which eigenvectors are
\[
\phi_1 + i \psi_1~, \forall \phi_1, \psi_1 \in \mathbb{O}.
\]

\subsection{The group   $GL(8,\mathbb{C})$}

In the previous subsections, we have presented two equivalent method to solve the octonionic eigenvalue problem. The translation of  octonionic left/right operators by $8\times 8$ real matrices, discussed in subsection 2.1, gives us the possibility to introduce a coupled eigenvalue equation characterized by {\em two} real numbers $a$ and $b$ and two real eigenvectors $\xi$ and $\eta$, see Eq.\,(\ref{coep}). This real coupled eigenvalue problem can be then translated by using the equivalence between the group $GL(8,\mathbb{R})$ and real linear  octonionic operators in a coupled real eigenvalue problem for octonionic operators,
\begin{eqnarray}
\label{oreal}
\mathcal{O}_\mathbb{R}\,\psi_\xi & = & a\, \psi_\xi - b\, \psi_\eta\,\,, \nonumber\\
\mathcal{O}_\mathbb{R}\,\psi_\eta & = & a\, \psi_\eta + b\, \psi_\xi\,\,,
\end{eqnarray}
with $\psi_\xi$ and $\psi_\eta$ $\in \mathbb{O}$, $a$ and $b$ $\in \mathbb{R}$ and $\mathcal{O}_\mathbb{R}$ represented by a real linear left/right octonionic operator. The presence of two real coupled eigenvalues suggests, see subsection 2.3,  the possibility to re-write such a coupled real eigenvalue problem as a ``complex" eigenvalue problem. Indeed,  by introducing a {\em new} complex imaginary unit $i$ which commutes with the imaginary units of the octonionic field $e_{_{1,2,...,7}}$, we find
\begin{equation}
\label{ocomp}
\mathcal{O}\,(\underbrace{\psi_\xi + i\, \psi_\eta}_{\Psi\,\,\in\,\, \mathbb{O}\,\times\, \mathbb{C}})  =  (\underbrace{a+i\,b}_{z\,\,\in\,\,\mathbb{C}})\, \left( \psi_\xi + i\, \psi_\eta \right)\,\,.
\end{equation}
The introduction of this complex imaginary unit also suggests to extend the matrix translation for  real linear octonionic operators,
\[\mathcal{O}_\mathbb{R} \,\,\,\,\, \leftrightarrow \,\,\,\,\, GL(8,\mathbb{R})\,\,,\]
introduced in ref.\cite{DK97} e re-presented in subsection 2.1, to {\em complex} linear octonionic operators
\[\mathcal{O}_\mathbb{C} \,\,\,\,\, \leftrightarrow \,\,\,\,\, GL(8,\mathbb{C})\,\,.\]
Thus, the two equivalent methods to solve the eigenvalue problem for a real linear octonionic operators, presented in subsections 2.2 and 2.3, can be used to solve the eigenvalue problem for complex linear octonionic operators. Observing that a complex linear octonionic operator, $\mathcal{O}_\mathbb{C}$, is characterized by two real linear octonionic operators ( $\mathcal{O}_{1\,\mathbb{R}}$ and $\mathcal{O}_{2\,\mathbb{R}}$ - real and complex  part of the complex linear octonionic operator $\mathcal{O}_\mathbb{C}$)  we can use the matrix translation
( $\mathcal{O}_{1\,\mathbb{R}} \to M_1$ and $\mathcal{O}_{2\,\mathbb{R}} \to M_2$)
  to represent the complex linear operator by  the $8\times8$ complex matrix $M:=M_1 + i\, M_2$. By solving the eigenvalue problem for the complex matrix $M$ we find its eigenvectors $\psi\in \mathbb{C}_{8\times 1}$ and eigenvalues $z\in \mathbb{C}$. Then, we translate back to the octonionic formalism
\begin{equation}
\mathcal{O}_\mathbb{C} \,\Psi = z\,\Psi\,\,,
\end{equation}
with $\Psi\in\mathbb{O}\times \mathbb{C}$ and $z\in \mathbb{C}$. After a discussion of the hermicity of octonionic operators and the appropriate use of scalar product to define probability amplitudes, we shall give an explicit example of a physical problem where this translation can be applied.

\section{On the hermiticity of octonionic matrices and operators}

An important step towards a generalization of standard quantum
theories is the use of complex scalar products or complex
geometry as referred to by Rembieli\'nski~\,\cite{R78}.

In quantum mechanics, it's well known that an anti-Hermitian
operator obeys the following rule
\begin{equation}
\label{defantiherm}
\hspace*{-2cm}\langle \psi, \mathcal{O} \varphi \rangle = -
\langle \mathcal{O} \psi , \varphi \rangle~~\Rightarrow~~\int \,
\psi^\dag ( \mathcal{O} \varphi) \, dx = - \int  \,( \mathcal{O}
\psi)^\dag \varphi \, dx = - \int \, ( \psi^\dag \mathcal{O}^\dag
) \varphi \, dx
\end{equation}
Nevertheless, while in complex and quaternionic quantum mechanics
we can define a direct correspondence between Hermitian matrices
and Hermitian operators, in octonionic quantum mechanics this is
not possible. For example, we shall see
that the matricial representation of $e_m$ is an
anti-Hermitian matrix, but no imaginary unit $e_m$ represents an
anti-Hermitian operator \,\cite{DK97}. In fact, given
$\psi:\mathbb{R} \to \mathbb{O}$ and $\varphi:\mathbb{R} \to
\mathbb{O}$, octonionic functions of a real variable $x$,
\[
\psi(x) = \psi_0(x) + \sum_{n=1}^7 \psi_n(x) \,
e_n~~~\mbox{and}~~~ \varphi(x) = \varphi_0(x) + \sum_{n=1}^7
\varphi_n(x) \, e_n~,
\]
where each $\psi_n$ and each $\varphi_n$ are real valued
functions, the nonassociativity of the octonionic algebra implies
that
\begin{equation}
\int \, \psi^\dag (e_m \varphi)\, dx = \langle \psi, e_m \varphi
\rangle \neq - \langle e_m \psi , \varphi \rangle = \int  \,
(\psi^\dag e_m) \varphi \, dx
\end{equation}
This contrasts with the situation within complex and quaternionic
quantum mechanics. Such a difficulty is overcome by using a
complex projection of the scalar product (complex geometry) with
respect to one of our imaginary units. We break the symmetry
between the seven imaginary units $e_1, \cdots , e_7$ and choose
as projection plane the one characterized by $(1, e_1)$. The new
scalar product is quickly obtained by performing, in the standard
definition, the following substitution:
\begin{equation}
\displaystyle \int F(x) \, dx ~~~\to~~~\displaystyle \int_c F(x) \, dx
\equiv \frac{1}{2} \left[ \displaystyle \int F(x) \, dx - e_1
\left(
 \displaystyle \int F(x) \, dx \right) e_1 \right]
\end{equation}
Working in Octonionic Quantum Mechanics with complex geometry,
$e_m$ represents an anti-Hermitian operator. In order to simplify
the proof for the imaginary unit $e_1$, we write the octonionic
functions $\psi$ and $\varphi$ as follows:
\begin{equation}
\begin{array}{l}
\psi = \tilde{\psi}_1 + e_2 \tilde{\psi}_2 + e_4 \tilde{\psi}_3 + e_6 \tilde{\psi}_4\\
\varphi = \tilde{\varphi}_1 + e_2 \tilde{\varphi}_2 + e_4
\tilde{\varphi}_3 + e_6 \tilde{\varphi}_4
\end{array}
\end{equation}
where $\tilde{\psi}_m$ and $\tilde{\varphi}_m$, $m=1,2,3,4$ are
complex-valued functions. The anti-hermiticity of $e_1$ is
recovered if
\begin{equation}
\label{exah} \int_c \psi^\dag(e_1 \varphi) dx = - \int_c (e_1
\psi)^\dag \varphi dx
\end{equation}
holds. Algebraic manipulation shows that, after the complex
projection, the only non vanishing terms are represented by {\it
diagonal} terms, that is, terms that contains the functions
$\tilde{\psi}^{\dag}_1 \tilde{\varphi}_1, \tilde{\psi}^{\dag}_2
\tilde{\varphi}_2 , \tilde{\psi}^{\dag}_3 \tilde{\varphi}_3$ and
$\tilde{\psi}^{\dag}_4 \tilde{\varphi}_4$. In fact, terms like
$\tilde{\psi}^{\dag}_1 e_5 \tilde{\varphi}_3$ or
$\tilde{\psi}^{\dag}_3 \tilde{\varphi}_4$ are annulled by complex
projection,
\begin{eqnarray}
\tilde{\psi}^{\dag}_1 (e_5 \tilde{\varphi}_3) = (\alpha_{_0} - e_1
\alpha_{_1})
(e_5 \beta_{_0} + e_4 \beta_{_1}) = \rho \,e_4 + \sigma \,e_5\\
(\tilde{\psi}^{\dag}_3 e_4) (- e_7 \tilde{\varphi}_4) = -
(\gamma_{_0} e_4 - \gamma_{_1} e_5)(e_7 \delta_{_0} - e_6
\delta_{_1}) = \varrho \, e_2 + \varsigma \, e_3
\end{eqnarray}
with $\alpha_{_{0,1}}, \beta_{_{0,1}}, \gamma_{_{0,1}},
\delta_{_{0,1}}, \rho, \sigma, \varrho$ e $\varsigma$ real
numbers. The diagonal terms give \\

\hspace*{-2mm}
$
\int_c \psi^\dag(e_1 \varphi) dx = \int_c \tilde{\psi}^{\dag}_1
(e_1 \tilde{\varphi}_1) - (\tilde{\psi}^{\dag}_2 e_2) [e_1 \, (e_2
\tilde{\varphi}_2)] - (\tilde{\psi}^{\dag}_3 e_4) [e_1 \, (e_4
\tilde{\varphi}_3)] - (\tilde{\psi}^{\dag}_4 e_6) [e_1 \, (e_6
\tilde{\varphi}_4)] dx
$\\

\hspace*{-5.5mm}
$
- \int_c ( e_1 \psi)^\dag \varphi dx = \int_c
(\tilde{\psi}^{\dag}_1 e_1) \tilde{\varphi}_1 -
[(\tilde{\psi}^{\dag}_2 e_2) \, e_1] (e_2 \tilde{\varphi}_2)] -
[(\tilde{\psi}^{\dag}_3 e_4)\, e_1]  (e_4 \tilde{\varphi}_3)] -
[(\tilde{\psi}^{\dag}_4 e_6) \, e_1 ]  (e_6 \tilde{\varphi}_4)] dx
$\\

\noindent
The parentheses above are not relevant since the first term
$\tilde{\psi}^{\dag}_1 e_1 \tilde{\varphi}_1$ is a complex number
and the three remaining terms respectively $\tilde{\psi}^{\dag}_2
e_2 e_1 e_2 \tilde{\varphi}_2$ (subalgebra 123),
$\tilde{\psi}^{\dag}_3 e_4 e_1 e_4 \tilde{\varphi}_3$ (subalgebra
145) and $\tilde{\psi}^{\dag}_4 e_6 e_1 e_6 \tilde{\varphi}_4$
(subalgebra 167), are quaternions. This proves that equation
(\ref{exah}) holds. Let us discuss the octonionic Hermitian
operator. Following the well known definition we have
$\mathcal{O}$ is an Hermitian operator if
\begin{equation}
\label{defherm} \langle \psi, \mathcal{O} \varphi \rangle =
\langle \mathcal{O} \psi , \varphi \rangle~~\Rightarrow~~\int
\psi^\dag ( \mathcal{O} \varphi) ~dx = \int( \mathcal{O}
\psi)^\dag \varphi ~dx = \int( \psi^\dag \mathcal{O}^\dag )
\varphi ~dx
\end{equation}

Now, suppose $\mathcal{O}$ an octonionic Hermitian operator and
\begin{equation}
\label{eigsys} \mathcal{O} \psi = \psi \lambda~,~~~\lambda \in
\mathbb{O}~.
\end{equation}
Applying the definition of hermiticity is easy to prove that
$\lambda$ must be real. In fact, let be $\psi = \varphi$, then
\begin{equation}
\langle \psi,  \mathcal{O} \psi \rangle = \langle \mathcal{O} \psi
, \psi \rangle~~~\Rightarrow~~~\psi^\dag ( \mathcal{O} \psi) = (
\mathcal{O} \psi)^\dag \psi = (\psi^\dag  \mathcal{O}^\dag) \psi
\end{equation}
which gives, after using equation (\ref{eigsys}),
\begin{equation}
\label{re}
 \psi^\dag ( \psi \lambda ) = ( \lambda^*  \psi^\dag)
\psi~.
\end{equation}
But octonionic numbers, $o_1, o_2 \in \mathbb{O}$, satisfy the
property
\begin{equation}
o_1^\dag ( o_1 o_2) = (o_1^\dag o_1) o_2 = (o_2 o_1^\dag) o_1~.
\end{equation}
So, returning to equation (\ref{re}) and using the property above
we have
\begin{equation}
\psi^\dag ( \psi \lambda ) = ( \psi^\dag  \psi ) \lambda =
|\psi|^2 \lambda
\end{equation}
and
\begin{equation}
( \lambda^\dag  \psi^\dag) \psi = \psi^\dag (\psi \lambda^\dag) =
(\psi^\dag \psi) \lambda^\dag = |\psi|^2 \lambda^\dag~.
\end{equation}
Since $|\psi|^2$ is real we obtain
\begin{equation}
\lambda = \lambda^\dag~~~\Rightarrow~~~\lambda \in \mathbb{R}.
\end{equation}
This means that even in the octonionic formalism, given an
Hermitian operator its eigenvalues must be real. Now, consider the
example given by Dray et al, in \,\cite{DJM00}. Given $M$ an
Hermitian matrix
\[
M = \left(  \begin{array}{rc}
          1 & e_4\\
          -e_4 & 1
        \end{array}  \right)
\]
we find
\begin{equation} \label{ex}
 \left(  \begin{array}{rc}
          1 & e_4\\
          -e_4 & 1
        \end{array}  \right) \, \left(  \begin{array}{c}
            e_5\\
            e_7
        \end{array}  \right) =  \left(  \begin{array}{c}
            e_5\\
            e_7
        \end{array}  \right) \, (1 - e_6)~.
\end{equation}
So, the Hermitian matrix $M$ has an eigenvalue given by $1 - e_6$
($\in \mathbb{O}$) and $( e_5~~e_7 )^t$ is an eigenvector
associated with it. Since the matrix is Hermitian we should expect
{\it real} eigenvalues but this doesn't happen. How can we prove
that the matrix $M$ does not represent an Hermitian operator?
Consider the example (\ref{ex}) and let be $\psi = \varphi$. So,
we have
\[
\psi^\dag ( \mathcal{O} \psi) = \psi^\dag ( \psi \lambda ) = (
-e_5~~~-e_7) \, \left[   \left(
 \begin{array}{c}
            e_5\\
            e_7
        \end{array}  \right) \, (1 - e_6) \right] =
\]
\begin{equation}
\hspace*{-2cm}
( -e_5~~~-e_7) \, \left(
\begin{array}{c}
            e_5 + e_3\\
            e_7 - e_1
        \end{array}  \right) = -e_5 (e_5 + e_3) - e_7 (e_7 - e_1)
        = 1 - e_6 + 1 - e_6 = 2 - 2\,e_6
\end{equation}
Now, calculating $(\mathcal{O} \psi)^\dag \psi$ we find
\[
(\mathcal{O} \psi)^\dag \psi = \left( \begin{array}{c}
            e_5 + e_3 \\
            e_7 - e_1
        \end{array}  \right)^\dag \left( \begin{array}{c}
            e_5\\
            e_7
        \end{array}  \right) = \left( \begin{array}{cc}
            -e_5 - e_3 &
            -e_7 + e_1
        \end{array}  \right) \left( \begin{array}{c}
            e_5\\
            e_7
        \end{array}  \right) =
\]
\begin{equation} - (e_5 + e_3)e_5 + (e_1 - e_7)e_7 =
        1 + e_6 + e_6 + 1 = 2 + 2\, e_6
\end{equation}
Thus, according to definition (\ref{defherm}) the operator
associated to the Hermitian matrix $M$ is not an Hermitian
operator since $\psi^\dag ( \mathcal{O} \psi) \neq (\psi^\dag \,
\mathcal{O}^\dag) \psi$.

At this point, we can assert that octonionic Hermitian matrices
are not associated with Hermitian operators so it's natural to
find non real eigenvalues for the matrices. We conclude that octonionic Hermitian
operators necessarily have real eigenvalues. Furthermore, a way to reobtain the
relation between Hermitian matrices and Hermitian operators is by
using the complex projection. It's worthy to observe that the
complex projection of the internal product gives the same result.
In the example just given
\begin{equation}
[\langle \psi,  \mathcal{O} \psi \rangle ]_\mathbb{C} =  [ \langle
\mathcal{O} \psi , \psi \rangle ]_\mathbb{C} = 2~.
\end{equation}

\section{Conclusions}

It is well known that amplitudes of probability have to be defined
in associative division algebras\,\cite{ADL}. Amplitudes of
probabilities defined in non-division algebras fail to satisfy the
requirement that, in the absence of quantum interference effects,
probability amplitude superposition should reduce to probability
superposition. The associative law of multiplication is needed to
satisfy the completeness formula. Amidst these constraints how can
be possible to formulate quantum theories by using wave functions
defined in non-division or non-associative algebras?  The answer
is very simple. The constraints concern the inner product and {\em
not} the Hilbert space in which we define our wave functions. The
amplitudes of probability have to be defined in $\mathbb{C}$ or
$\mathbb{H}$ but the vectors in the Hilbert space have {\em no}
constraint. The choice of quaternionic inner product seems to be
best adapted to investigate deviations from the standard complex
theory in quantum mechanics and quantum fields\,\cite{ADL}.

In this paper, we have suggested a solution for the octonionic
eigenvalue problem by a formulation based on  a coupled equation
with two real parameters which play the  same role of the complex
eigenvalue  in complex and quaternionic quantum mechanics. The
complex eigenvalue problem in complex and quaternionic quantum
mechanic can be obviously solved by a real coupled problem and this
should represent the complex and quaternionic limit of the
octonionic eigenvalue problem proposed in this article.

An intriguing result obtained from our investigation was that
octonionic Hermitian matrices do {\em not} necessarily represent
octonionic Hermitian operators. This is essentially due to the
fact  that  octonionic Hermitian matrices can have {\em not} real
eigenvalues. To overcome this problem, we need to introduce a
complex geometry, i.e. complex inner products. The natural choice is represented by the use of the imaginary unit $i$ which commutes with the octonionic imaginary units $e_{_{1,2,...,7}}$.

An interesting application of the material presented in this work can be immediately found in the octonionic formulation of the Dirac equation. The Dirac Hamiltonian\,\cite{ZUB},
\begin{equation}
\mathcal{H} = \boldsymbol{\alpha} \cdot  \boldsymbol{p}\,c + \beta\,m\,c^{^{2}}\,\,,\,\,\,\,\,\boldsymbol{p}=-\,i\,\hbar\,\nabla\,\,,
\end{equation}
which describes the temporal behavior of relativistic particles is given in terms of $4 \times 4$ complex matrices, $ \boldsymbol{\alpha}$ and $\beta$, satisfying the Dirac algebra
\[\{\, \boldsymbol{\alpha}\,,\,\beta \,\} =0\,\,\,,\,\,\,\,\,\{\,\alpha_m\,,\,\alpha_n\,\}=0\,\,\,\,\,\mbox{for}\,\, m\neq n\,\,\,,\,\,\,\,\,
\alpha_n^{^2}=\beta^{^{2}}=\mbox{I}_{_{4\times 4}}\,\,.\]
By using the quaternionic sub-algebra $e_{_{1,2,3}}$, we can immediately found a complexified quaternionic representation for the $\boldsymbol{\alpha}$ matrices,
\begin{equation}
\boldsymbol{\alpha}=i\,\boldsymbol{e}=\left(\,i\,e_{_{1}}\,,\,i\,e_{_{2}}\,,\,i\,e_{_{3}}\,\right)\,\,.
\end{equation}
The matrix $\beta$ can be represented by using, for example, the octonionic imaginary unit $e_{_{4}}$,
\begin{equation}
\beta = i\,e_{_{4}}\,\,.
\end{equation}
An octonionic representation for the Dirac Hamiltonion is thus given by
\begin{equation}
\mathcal{H} \,\,\,\to\,\,\,\hbar\,c\,\boldsymbol{e}\cdot \nabla  + i\,m\,c^{^{2}}e_{_{4}}\,\,.
\end{equation}
The complexified octonionic solution
\[ \underbrace{\psi_{_{0}}\,\,+\,\,\boldsymbol{e}\cdot\boldsymbol{\psi}}_{
\Psi\,\,\,\in\,\,\,\mathbb{H}\,\times\,\mathbb{C}} +\, e_{_{4}}\,(\,  \underbrace{\phi_{_{0}}+\boldsymbol{e}\cdot\boldsymbol{\phi}}_{\Phi\,\,\,\in\,\,\,
\mathbb{H}\,\times\,\mathbb{C}}\,)\,\,\,,\,\,\,\,\,
\psi_{_{0}},\,\boldsymbol{\psi},\,\phi_{_{0}},\,\boldsymbol{\phi}\,\,\,\in \mathbb{C}(1,i)\,\,,
\]
by using $\mathbb{C}$ inner products contains in this formulation two orthogonal spinorial solutions, $\Psi$ and $\Phi$, each of one with its $4$ complex degrees of freedom represent a Dirac particle. This suggests a natural e simple one dimensional octonionic formulation of the standard model, where two orthogonal spinorial solutions are needed to represent the leptonic and quark doublets\,\cite{SM}.

\end{document}